\documentstyle[12pt,epsf,graphics]{article}
\newcommand{\be}{\begin{equation}}
\newcommand{\ee}{\end{equation}}
\begin{document}
\begin{center}
{\bf\Large The shortest cut in  brane cosmology}
\end{center}
\vspace{1ex}
\centerline{\large
Elcio Abdalla$^1$, Bertha Cuadros-Melgar$^1$, Sze-Shiang
  Feng$^{1,2}$, Bin Wang$^{ 3}$}
\begin{center}
{\large 
$^1$ Instituto de Fisica, Universidade de Sao Paulo,
  C.P.66.318, CEP 05315-970, \\ Sao Paulo, Brazil \\ $^2$ Department of
  Astronomy and Applied Physics, University of Science and \\ Technology
  of China, 230026, Hefei, China \\ $^3$ Theory Division, Department of
  Physics, Fudan University \\ Shanghai 200433, P.R. China}
\end{center}
\begin{abstract}
We consider brane cosmology studying the shortest null path on the brane
for photons, and in the bulk for gravitons. We derive the differential
equation for the shortest path in the bulk for a 1+4 cosmological
metric. The time cost and the redshifts for 
photons and gravitons after traveling their respective path are
compared. We consider some numerical solutions of the shortest path
equation, and show that there is no shortest path in the bulk for the
Randall-Sundrum vacuum brane solution, the linear cosmological
solution of Bin\'etruy, {\it et al} for $\omega = -1, -\frac {2}{3}$,
and for some expanding brane universes.
\end{abstract}

{PACS numbers:75.20.Hr,71.10.Hf,71.27.+a,71.55.-i}
\newpage
%%%%%%%%%%%%%%%%%%%%%%%%%%%%%%%%%%%%%%%%%%%%%%%%%%%%%%%%%%%%%%%%%%%%%%%%%%%%%
\section{ Introduction} 
%%%%%%%%%%%%%%%%%%%%%%%%%%%%%%%%%%%%%%%%%%%%%%%%%%%%%%%%%%%%%%%%%%%%%%%%%%%%%
The possibility of using extra dimensions in order to explain features
related to unified field theories has been advocated several decades ago
by Kaluza and Klein. After a die out for many years such an idea was
reestablished in the context of supergravity and string theory, especially
in the latter, where extra dimensions are required in order that
the theory is rendered well defined. Meanwhile other problems have been
posed in the framework of unified theories. One of them is the huge hierarchy
between the electro-weak scale ($\sim 100$ GeV) and the Planck scale
($\sim 10^{19}$ TeV). One possibility to explain that difference is based
on the dynamics of supersymmetry, a very beautiful idea that has not,
unfortunately, rendered due (and ripe) issues. In the usual Kaluza-Klein,
and also in the modern proposals to deal with extra dimensions, while the
1+3 (physical) dimensions open up to infinity, the extra dimensions are
confined in a region of the size of the Planck length, namely $\sim
10^{-33}$cm, staying beyond experimental verification, today or in the
near future. 

However, it has been recently shown that it is possible to explain the
hierarchy between the electro-weak  and the Planck scale by dimensional
reduction without compactifying the extra dimensions. Moreover, the usual 1+3
dimensional Einstein theory of gravity can be reproduced on the
macroscopic distance scale \cite{s1}-\cite{s5}. This is quite different
from the standard approach, in which extra dimensions open up at short
distances only, whereas above a certain length scale, physics is
effectively described by 1+3 dimensional theories. Our 1+3 dimensional
Universe would be a three dimensional brane living in a higher dimensional
theory, thus displaying a certain number of additional dimensions.
A further proposal to deal with the additional dimensions is to have them
compactified in a submilimeter scale, unifying in a natural way the
electro weak and Planck scales \cite{add}.

The possibility of relaxing the constraints on the size of the extra
dimensions is very appealing. Such is the case of the Randall-Sundrum
(RS) model \cite{s1,s2}, where the Universe is 1+4 dimensional and the Standard
Model fields are localized on a 3-brane embedded in the 4-dimensional
space. Only gravitational fields can propagate in all four space
directions. At the phenomenological length scale the Kaluza-Klein
zero-modes are responsible for the well-posed Einstein 1+3 dimensional
theory of gravity and the excitations provide a correction. Due to the
"warp factor" of the brane, a mass scale around that of Planck mass
corresponds to a TeV mass scale in the {\it visible} brane. This explains
the hierarchy problem. The cosmological consequence of this model is also
under active investigation \cite{s6}-\cite{s12}. The model leads to
new perspectives in many interesting aspects such as the question of the
cosmological constant. 

The construction of the brane-universe can be traced to the study of
$E_8\times E_8$ string theory, presumably 11-dimensional, with the field
theory limit studied in \cite{horava-witten}, and where matter fields live
in 10-dimensional branes at the edge of the space-time. The issue of
higher dimensionality and its consequences for the early universe have
been often discussed in the recent literature \cite{inflation}. 
Problems related to higher derivative gravity \cite{so1} and on the
cosmological constant problem \cite{so2} have also been studied, besides
the AdS/CFT correspondence and Cardy formula \cite{bwabdsu}.

\begin{figure}[htb]
\begin{center}
\leavevmode
\begin{eqnarray}
\epsfxsize= 6truecm%\rotatebox{}
{\epsfbox{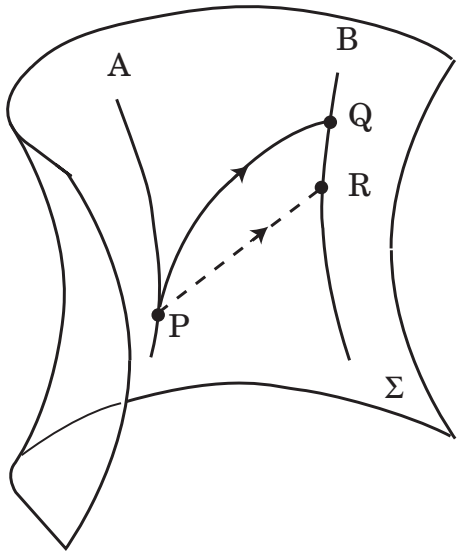}}\nonumber
\end{eqnarray}
\caption{Possible two paths for massless signal
  propagation. Solid curve PQ is a null geodesic on the brane $\Sigma$
  and broken line PR is a null geodesic in the bulk (modified from \cite{s13})
  \label{fig1}
.}
\end{center}
\end{figure}

In spite of the attractive aspects of the model, causality can
be violated, as first noticed in \cite{s13} and \cite{s14}. We have two
choices facing this situation. Either we accept the viewpoint that
true causality should be defined by the null geodesics in the 1+4
universe instead of in the 1+3 brane spacetime or we find some
mechanism to avoid such a violation on the brane. In the first case,
the violation must be neglectable in low energy experiments,
otherwise, it could have been already found. The question is whether 
it could be substantial in cosmology. If the answer is positive, it
might help solving the well known horizon problem as discussed in
\cite{s13} and \cite{s14}. In this paper, we consider the following
problem. Suppose there are two observers $A$ and $B$ on the brane. $A$ can
send series of photons or gravitons to $B$ in order to establish
communication  (see Fig. \ref{fig1}). According to the brane
cosmology, photons travel on the brane while gravitons may travel in the
bulk. We consider the three questions: (i) what is the shortest path for
gravitons, and whether it is on the brane or in the bulk; (ii) how earlier the
gravitons can arrive at $B$; (iii) what is the difference of the
redshift for photons and gravitons after they arrive at $B$.
%%%%%%%%%%%%%%%%%%%%%%%%%%%%%%%%%%%%%%%%%%%%%%%%%%%%%%%%%%%%%%%%%
%%%%%%%%%%%%%%%%%%%%%%%%%%%%%%%%%%%%%%%%%%%%%%%%%%%%%%%%%%%%%%%%%%%%%%%%%
%%%%%%%%%%%%%%%%%%%%%%%%%%%%%%%%%%%%%%%%%%%%%%%%%%%%%%%%%%%%%%%%%%%%%%%%%%%%%
\section{Preliminaries}
%%%%%%%%%%%%%%%%%%%%%%%%%%%%%%%%%%%%%%%%%%%%%%%%%%%%%%%%%%%%%%%%%%%%%%%%%%%%%
We shall consider a 5-dimensional  metric describing brane cosmology.
We thus set up a 5-dimensional action of the form \cite{s7}
\be
S^{(5)}=-\frac1{2\kappa_5^2}\int d^5x \sqrt{-\tilde g}\tilde R +\int d^5x
\sqrt{-\tilde g} {\cal L}_m\quad .
\ee
The constant $\kappa_5$ is related to the Planck mass as
$\kappa_5^2=M_{Pl}^{-3}$. The 5-dimensional metric is
\begin{equation}\label{metric}
ds^2 _5 = -n^2(t,y) dt^2 + a^2(t,y) \gamma_{kj} dx^k dx^j + b^2(\tau,y)dy^2
\end{equation}
where $\gamma_{kj}$ represents a maximally symmetric 3-metric. The
energy-momentum appearing in the Einstein equation $G_{AB} =\kappa^2
_5 {\cal T}_{AB}$ is decomposed as
\begin{equation}
{\cal T}_{AB} = \hat T_{AB} + T_{AB}
\end{equation}
where $\hat T_{AB}$ is the energy-momentum tensor of the bulk matter
(in the RS scenario it comes from the bulk cosmological constant
$\Lambda$, that is, $\hat T_{AB} = -\Lambda \delta ^A _B$) and $T_{AB}$
corresponds to the matter content on the brane located at $y=0$. We are
interested in the case where the energy-momentum tensor of the bulk matter
can be expressed as 
\begin{equation}
T^A _B = \frac {\delta(y)} {b} diag (-\rho-\sigma, p-\sigma,
p-\sigma, p-\sigma, -\sigma)\quad .
\end{equation}
Here, $\sigma$ is the brane tension in the RS scenario. The
energy-density $\rho$ and the pressure $p$ come from the ordinary matter
on the brane and are independent of the position. Assuming the ${\bf
  Z}_2$-symmetry and $\sigma=0$, the Einstein equation permits the
following exact cosmological brane solution \cite{s7} (corresponding
to $\Lambda=0$, $\sigma=0$, $\gamma_{jk}=\delta_{jk}$) 
\begin{eqnarray}
a &=& a_0(t) (1+\lambda |y|) \nonumber \\
n &=& n_0(t) (1+\mu |y|) \\\label{cosmosol}
b &=& b_0 \nonumber
\end{eqnarray}
where $b_0$ is constant in time (a redefinition of $y$ renders it to
be 1) and $n_0(t)$ is an arbitrary function (a suitable redefinition of $t$
fixes it to be 1). In the above, 
\begin{equation}
\lambda=- {{\kappa^2 _5} \over 6} b_0 \rho
\end{equation}
\begin{equation}\label{mu}
\mu={{\kappa^2 _5} \over 2} \left(\omega + {2 \over 3} \right) b_0 \rho
\end{equation}
where $\kappa^2 _5$ is related to the 5-dimensional Newton constant
$G_5$ by $\kappa^2 _5 = 8 \pi G_5$, and the matter equation of state is
$p=\omega \rho$ as usually.

For $\omega=-1$ we have the inflationary case,
\begin{equation}
a_0(t) = e^{Ht}, \quad H={\kappa^2 \over 6} \rho =const.,
\end{equation}
while for $\omega \neq -1$, the usual solution arises,
\begin{equation}
a_0=t^q, \quad \kappa^2 _5 \rho = {{6q} \over t}, \quad q={1 \over
  {3(1+\omega)}}. 
\end{equation}
Remarkably, the exact solution in the RS model can also be obtained
\cite{s15}. Note that the parameters $\rho_b$ and $p_b$ in \cite{s15}
are related to the corresponding ones here in this paper by the
relations $\rho_b =\rho+\sigma$, $p_b=p-\sigma$. The solution can be
written in terms of the function 
\begin{eqnarray}
a(t,y) &=& \left\{ {1 \over 2} \left( 1+ {{\kappa^2 _5
        (\sigma+\rho)^2} \over {6\Lambda}} \right) a^2 _0 + {{3{\cal
        C}} \over {\kappa^2 _5 \Lambda a^2 _0}} \right. \nonumber \\
&& \left. + \left[ {1 \over 2} \left(1- {{\kappa^2 _5
        (\sigma+\rho)^2} \over {6\Lambda}} \right) a^2 _0 - {{3{\cal
        C}} \over {\kappa^2 _5 \Lambda a^2 _0}} \right] \cosh(\mu y)
        \right. \nonumber \\
&& \left. - {{\kappa_5 (\sigma+\rho)} \over {\sqrt{-6\Lambda}}} a^2 _0
        \sinh (\mu |y|)\right\}^{1/2}\quad .
\end{eqnarray}

We now construct the remaining function
\begin{equation}
n(t,y) = {{\dot a(t,y)} \over {\dot a_0(t)}} \quad .
\end{equation}
As for eq. (33) in \cite{s15}, we also have
\begin{equation}
\dot \rho + 3 {{\dot a_0} \over a_0} (\rho +p) =0\quad .
\end{equation}
Defining
\begin{equation}
\lambda=\sqrt{ {\Lambda \over {6\kappa^2 _5}} + {\sigma^2 \over
    {36}}}\quad ,
\end{equation}
and assuming $\lambda \geq 0$ and $p=\omega \rho$, the Friedman
equation can be solved in the case ${\cal C}=0$, $k=0$. For $\lambda >
0$,
\begin{equation}
a_0(t) = a_\star \rho^q _\star \left\{ {\sigma \over {36 \lambda^2}}
  \left[ \cosh (\kappa^2 _5 \lambda t/q) -1 \right] + {1 \over
  {6\lambda}} \sinh (\kappa^2 _5 \lambda t/q) \right\} ^q .
\end{equation}
For $\lambda =0$, which is the case of RS model,
\begin{equation}
a_0(t) = a_\star (\kappa^2 _5 \rho_\star)^q \left( {1 \over {72q^2}}
  \kappa^2 _5 \sigma t^2 + {1 \over {6q}}t \right)^q
\end{equation}
where $a_\star$, $\rho_\star$ are constant (the origin of time being
chosen so that $a_0(0)=0$).

%%%%%%%%%%%%%%%%%%%%%%%%%%%%%%%%%%%%%%%%%%%%%%%%%%%%%%%%%%%%%%%%%%%%%%%%%%%%%
\section{ The shortest cut and the redshift}
%%%%%%%%%%%%%%%%%%%%%%%%%%%%%%%%%%%%%%%%%%%%%%%%%%%%%%%%%%%%%%%%%%%%%%%%%%%%%
\noindent{\it Equation for the shortest cut}.
 
We consider the generic metric (\ref{metric}) for $b=1$. Consider two
points, $r_A$ and $r_B$ on the brane. In general, there are more than
one null geodesic connecting $r_A$ to $r_B$ in the 1+4 spacetime. The
trajectories of photons must be on the brane and those of gravitons
may be outside as assumed here. We consider the
shortest path for both photons and gravitons. Since the 3-metric is
spherically symmetric, we can omit the angular part and just consider
the problem for
\begin{equation}
ds_3^2=-n^2(t,y)dt^2+a^2(t,y)f^2(r)dr^2+dy^2
\end{equation}
The photon path is on the brane ($n(t,0)=1$), therefore
\begin{equation}
-dt^2+a_0^2(t)f^2(r)dr^2=0,
\end{equation}
which can be immediately integrated as
\begin{equation}
\int^{r}_{r_A}f(r^\prime)dr^\prime=\int^{t}_{t_A}
\frac{dt^\prime}{a_0(t^\prime)}\quad .
\end{equation}
The graviton path is defined in terms of the geodesic equation
\begin{equation}
-n^2(t,y)dt^2+a^2(t,y)f^2(r)dr^2+dy^2=0\quad .
\end{equation}

We suppose that the path is parameterized by $y=y(t)$. Thus the relation
$r=r(t)$ is obtained by
\begin{equation}\label{rsvac}
\int^{r}_{r_A}f(r^\prime)dr^\prime= \int^t_{t_A}
\frac{\sqrt{n^2(t,y)-\dot{y}^2(t)}}{a(t,y)}dt \quad .
\end{equation}
We are looking for the path for which, $t_B$ reaches its minimum when
$r=r_B$. For this purpose, we consider the general case
\begin{equation}
\int^{r_B}_{r_A}f(r^\prime)dr^\prime= \int^{t_B}_{t_A}
{\cal L}[y(t),\dot{y}(t);t]dt\quad .
\end{equation}
For an adjacent path $y=y(t)+\delta y(t)$, we have
\begin{equation}
\int^{r_B}_{r_A}f(r^\prime)dr^\prime= \int^{t_B+\delta t_B}_{t_A}
{\cal L}[y(t)+\delta y(t),\dot{y}(t)+\delta\dot{y}(t);t]dt
\end{equation}
therefore we find the usual condition
\begin{equation}
-\delta t_B{\cal L}[y(t_B),\dot{y}(t_B);t_B]=\delta
\int^{t_B}_{t_A}{\cal L}[y(t),\dot{y}(t);t]dt\quad .
\end{equation} 
The problem is transformed into the Euler-Lagrange problem
\begin{equation}
\delta\int^{t_B}_{t_A}{\cal L}[y(t),\dot{y}(t);t]dt =0\quad .
\end{equation}
In our case, 
\begin{equation}
{\cal L}[y(t),\dot{y}(t);t]=
\frac{\sqrt{n^2(t,y)-\dot{y}^2(t)}}{a(t,y)}\quad ,
\end{equation}
and we have
\begin{eqnarray}
\frac{\partial{\cal L}}{\partial y}&=&
-a^{-2}a^\prime(n^2-\dot{y}^2)^{1/2}+a^{-1}(n^2-\dot{y}^2)^{-1/2}nn^\prime
\nonumber \\ 
\frac{\partial{\cal L}}{\partial \dot{y}}&=&
-a^{-1}(n^2-\dot{y}^2)^{-1/2}\dot{y}\quad . 
\end{eqnarray}
The Euler-Lagrange equation thus reads
\begin{eqnarray}\label{e-l}
-\ddot{y}&+&(\frac{\dot{a}}{a}+\frac{\dot{n}}{n})\dot{y}
+(\frac{2n^\prime}{n}-\frac{a^\prime}{a})\dot{y}^2-\frac{\dot{a}}{an^2}
\dot{y}^3\nonumber\\
&&+(\frac{a^\prime}{a}n^2-nn^\prime)=0\quad .
\end{eqnarray}
From this equation we can see that the shortest path is on the brane only
when
\begin{equation}
\frac{a^\prime}{a}n^2-nn^\prime=0\quad ,
\end{equation}
i.e.
\begin{equation}
\partial_y(\frac{a}{n})=0\quad .
\end{equation}

Further, if there exists a solution, when $y$ reaches its maximum,
where $\dot{y}=0$ and $\ddot{y}<0$, we have
\begin{eqnarray}
-\ddot{y}+(\frac{a^\prime}{a}n^2-nn^\prime)=0\quad .
\end{eqnarray}
Thus, $\frac{a^\prime}{a}n^2-nn^\prime$, i.e. $\partial_y(an^{-1})$ must be
negative at this point.

The equation is a very difficult nonlinear ordinary differential
equation. There is no guarantee for the existence of the 
required solutions. In order to obtain a solution with both two ends on the
brane, we can make the Fourier expansion
\begin{eqnarray}
y(t)&=&\sum^{+\infty}_{l=1}y_l\sin\frac{l\pi}{t_{gB}-t_A}(t-t_A) \quad ,\\
a(t,y)&=&A(y)+\sum^{+\infty}_{l=1}[a^s_l(y)\sin\frac{l\pi}{t_{gB}-t_A}(t-t_A)
\nonumber\\ 
 &&+a^c_l(y)\cos\frac{l\pi}{t_{gB}-t_A}(t-t_A)] \quad ,\\
n(t,y)&=&N(y)+\sum^{+\infty}_{l=1}[n^s_l(y)\sin\frac{l\pi}{t_{gB}-t_A}(t-t_A)
\nonumber\\ 
&&+n^c_l(y)\cos\frac{l\pi}{t_{gB}-t_A}(t-t_A)] \quad ,
\end{eqnarray}
and then substitute back into the differential equation to obtain the
coefficients $y_l$. 
Here $t_{gB}$ is the time when the graviton arrives at $r_B$, which is
different from the time $t_{\gamma B}$ when the photon arrives at
$r_B$. It should be determined self-consistently by the equation
\begin{equation}
\int^{r_B}_{r_A}f(r^\prime)dr^\prime= \int^{t_{gB}}_{t_A}
\frac{\sqrt{n^2(t,y)-\dot{y}^2(t)}}{a(t,y)}dt
\end{equation}
once the solution is obtained.

If we want to find the path for a graviton so that it can reach the
farthest within a given time interval $[t_A,t_B]$, we can also use the
Euler-Lagrange equation. Then the length difference between geodesics
for photons and gravitons within a given time interval can be evaluated 
\begin{equation}
\int^{r_g}_{r_A}f(r^\prime)dr^\prime= \int^{t_{B}}_{t_A}
\frac{\sqrt{n^2(t,y)-\dot{y}^2(t)}}{a(t,y)}dt
\end{equation}
\begin{equation}
\int^{r_{\gamma}}_{r_A}f(r^\prime)dr^\prime=\int^{t_B}_{t_A}\frac{dt^\prime}
{a_0(t\prime)}
\end{equation}

\noindent{\it Photon and graviton redshift}. 

In general, if $A$ sends out massless signals at $x^\mu _A$ and
 $x^{\mu}_A+dx^{\mu}_A$, these signals will reach $B$ at $x^\mu _B$
 and $x^{\mu}_B+dx^{\mu}_B$. The relation of $x^\mu _A$,
 $x^{\mu}_A+dx^{\mu}_A$ and $x^\mu _B$, $x^{\mu}_B+dx^{\mu}_B$ can be
 obtained by solving the geodesic equation. Then the redshift of
 the signal is \cite{s16}
\begin{equation}
\frac{\nu_B}{\nu_A}=\sqrt{\frac{g_{00}(x_B)}{g_{00}(x_A)}}
\frac{g_{0\mu}(x_A)dx^{\mu}_A}{g_{0\nu}(x_B)dx^{\nu}_B}
=\sqrt{\frac{g_{00}(x_A)}{g_{00}(x_B)}}\frac{dx^0_A}{dx^0_B}
\end{equation}
For a static metric such as the Schwarzschild case,
it can be shown that $dx^0_A=dx^0_B$, therefore,
\begin{equation}
\frac{\nu_B}{\nu_A}=\sqrt{\frac{g_{00}(x_A)}{g_{00}(x_B)}}\quad .
\end{equation}
For the time-dependent RW metric  we have
\begin{equation}
\frac{dx^0_A}{dx^0_B}=\frac{R(x^0_A)}{R(x^0_B)}\quad ,
\end{equation}
in which case the redshift is given by
\begin{equation}
\frac{\nu_B}{\nu_A}=\frac{R(x^0_A)}{R(x^0_B)}\quad .
\end{equation}
Thus, in the geometric-optics limit, the redshifts in the two cases
can be systematically discussed.

Here, we consider that another graviton starts traveling from
$r_A$ at a later time $t_A+\delta t_A$. Its shortest path is in general
different from the previous one. Let us denote it as $y_*=y_*(t)$. Then
the time when it arrives at $r_B$ will be a later time $t_{gB}+\delta t_{gB}$
\begin{equation}
\int^{r_B}_{r_A}f(r^\prime)dr^\prime= \int^{t_{gB}+\delta
t_{gB}}_{t_A+\delta t_A}
\frac{\sqrt{n^2(t,y_*)-\dot{y_*}^2(t)}}{a(t,y_*)}dt \quad .
\end{equation}
Therefore we have the equality
\begin{equation}
\int^{t_{gB}}_{t_A}
\frac{\sqrt{n^2(t,y)-\dot{y}^2(t)}}{a(t,y)}dt
=\int^{t_{gB}+\delta
t_{gB}}_{t_A+\delta t_A}
\frac{\sqrt{n^2(t,y_*)-\dot{y_*}^2(t)}}{a(t,y_*)}dt  \quad .
\end{equation}
For infinitesimal $dt_A$ and $dt_B$, we have
\begin{equation}
dt_B
\left(\frac{\sqrt{n^2(t,y)-\dot{y}^2(t)}}{a(t,y)}\right)\Bigl\arrowvert
  _B=
dt_A
\left(\frac{\sqrt{n^2(t,y)-\dot{y}^2(t)}}{a(t,y)}\right)\Bigl\arrowvert
_{A}
\end{equation}
Thus, the graviton redshift is given by
\begin{equation}
\frac{\nu_{gB}}{\nu_{gA}}=\frac{a_0(t_A)}{a_0(t_B)}\sqrt{\frac{1-\dot{y}^2(t_B)
}{1-\dot{y}^2(t_A)}} \quad , 
\end{equation}
while for the photon we have
\begin{equation}
\frac{\nu_{gB}}{\nu_{gA}}=\frac{a_0(t_A)}{a_0(t_B)} \quad .
\end{equation}

%%%%%%%%%%%%%%%%%%%%%%%%%%%%%%%%%%%%%%%%%%%%%%%%%%%%%%%%%%%%%%%%%%%%%%%%%%%%%
\section{ Examples}
%%%%%%%%%%%%%%%%%%%%%%%%%%%%%%%%%%%%%%%%%%%%%%%%%%%%%%%%%%%%%%%%%%%%%%%%%%%%%
\noindent {\it RS vacuum solution}\cite{s1} \cite{s2}. 

In this case 
\begin{equation}
n(y,t)=a(y,t)=e^{-k|y|} \quad .
\end{equation}
Eq. (\ref{e-l}) turns out to be
\begin{equation}
\ddot{y}+k{y}^2=0
\end{equation}
It has two possible solutions, one is $y=y_A=0$, and the other is
$y=y_0+k\ln(t-t_0)$. The second solution does not meet our requirement
because it will not end on the brane. So the shortest path must be on the
brane. This agrees with the conclusion in \cite{s13}.  

\noindent {\it The linear cosmological solution}. 

We first consider the case $\omega=-\frac{2}{3}$ so that from
(\ref{mu}) $\mu=0$, $a(t,y)=t-y$, $\lambda=-{1 \over t}$. The
equation is
\begin{equation}
-(t-y)\ddot{y}+\dot{y}+\dot{y}^2-\dot{y}^3-1=0
\end{equation}
Let $t-y=u$, then
\begin{equation}
u\ddot{u}+\dot{u}^3-2\dot{u}^2=0 \quad ,
\end{equation}
or
\begin{equation}
\frac{1}{2\dot{u}^2-\dot{u}^3}\frac{d}{dt}\dot{u}^2=\frac{2\dot{u}}{u} \quad .
\end{equation}
Therefore,
\begin{eqnarray}
\int\frac{d\dot{u}}{2\dot{u}-\dot{u}^2}&=&\int\frac{du}{u} \quad ,\\
\frac{\dot{u}}{2-\dot{u}}&=&cu^2\quad .
\end{eqnarray}
We can obtain the solution ($t_0$ and $c$ are two integration constants)
%\begin{equation}
%2(t-t_0)=u-\frac{1}{cu} \quad .
%\end{equation}
%We thus obtain the solution
\begin{equation}
y=t\pm\sqrt{(t-t_0)^2+\frac{1}{c}} \quad .
\end{equation}
It is obvious that this path can not end on the brane either. 
Furthermore, we consider the case $\omega=-1$, $\lambda=\mu=const.\,
a_0(t)=e^{Ht}$. So $\partial_y(a/n)=0$. Therefore the shortest path
is on the brane.

\noindent {\it The general linear cosmological solution} \cite{s7}. 

Consider the case $\omega\not=-1$  
\begin{eqnarray}
a_0(t)&=&t^q,\qquad \lambda=-\frac{q}{t}\quad ,\qquad 
\mu =w\frac{q}{t}\quad ,\qquad  w=2+3\omega\\
a(t,y)&=&t^q-qt^{q-1}y  \quad ,\qquad 
n(t,y)=1+\frac{q\omega}{t}y\\
\dot{a}(t,y)&=&qt^{q-1}-q(q-1)t^{q-2}y  \quad ,\qquad 
a^\prime(t,y)=-qt^{q-1} \\
\dot{n}(t,y)&=&-q\omega t^{-2}y  \quad ,\qquad 
n^\prime(t,y)=q\omega t^{-1}
\end{eqnarray}

%%%%%%%%%%%%%%%%%%%%%% Begin Figure 2 %%%%%%%%%%%%%%%%%%%%%%%%%%%%%%
\begin{figure}[t]%
\begin{center}
\leavevmode       
\begin{eqnarray}
\epsfxsize= 8truecm\rotatebox{-90}
{\epsfbox{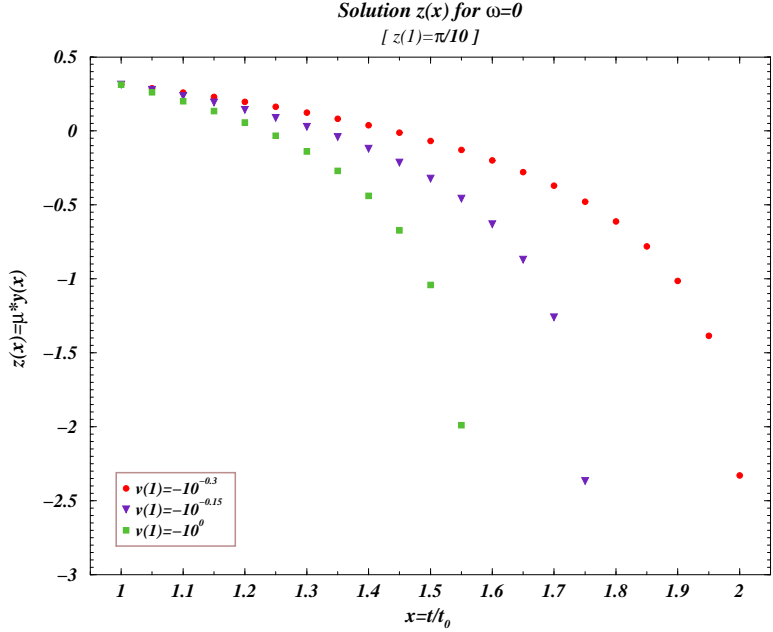}}  \nonumber
\end{eqnarray}
\end{center}
\caption{{Diagram for $y \sim 0.3 \ell_P$.}} \label{2}
\end{figure}
%%%%%%%%%%%%%%%%%%%%%%% End Figure 2  %%%%%%%%%%%%%%%%%%%%%%%%%%%%%% 
%%%%%%%%%%%%%%%%%%%%%% Begin Figure 3 %%%%%%%%%%%%%%%%%%%%%%%%%%%%%%
\begin{figure}[t]%[bh]
\begin{center}
\leavevmode       
\begin{eqnarray}
\epsfxsize= 8truecm\rotatebox{-90}
{\epsfbox{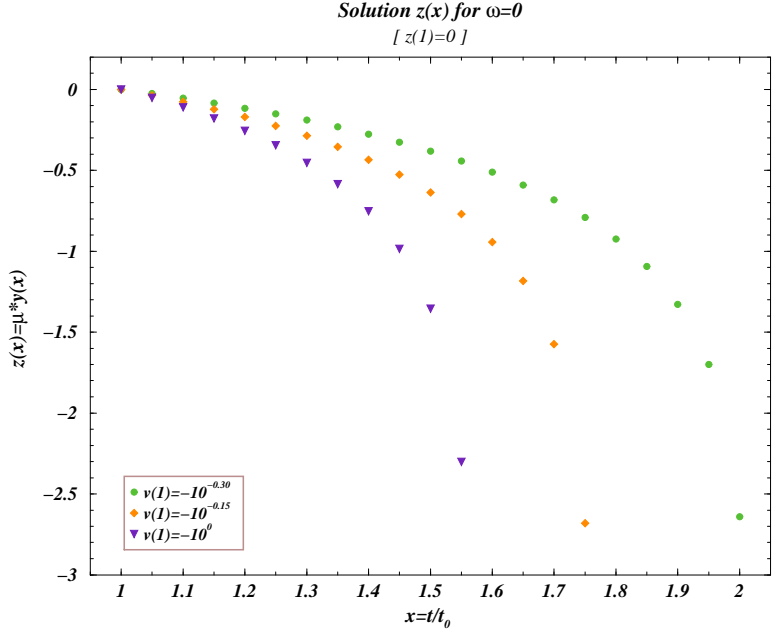}}  \nonumber
\end{eqnarray}
\end{center}
\caption{{The same diagram as before, with $y$ beginning at 
    the brane.}}
\label{3}
\end{figure}
%%%%%%%%%%%%%%%%%%%%%%% End Figure 3  %%%%%%%%%%%%%%%%%%%%%%%%%%%%%% 

%%%%%%%%%%%%%%%%%%%%%% Begin Figure 4 %%%%%%%%%%%%%%%%%%%%%%%%%%%%%%
\begin{figure}[t]%%[ht]
\begin{center}
\leavevmode       
\begin{eqnarray}
\epsfxsize= 8truecm\rotatebox{-90}
{\epsfbox{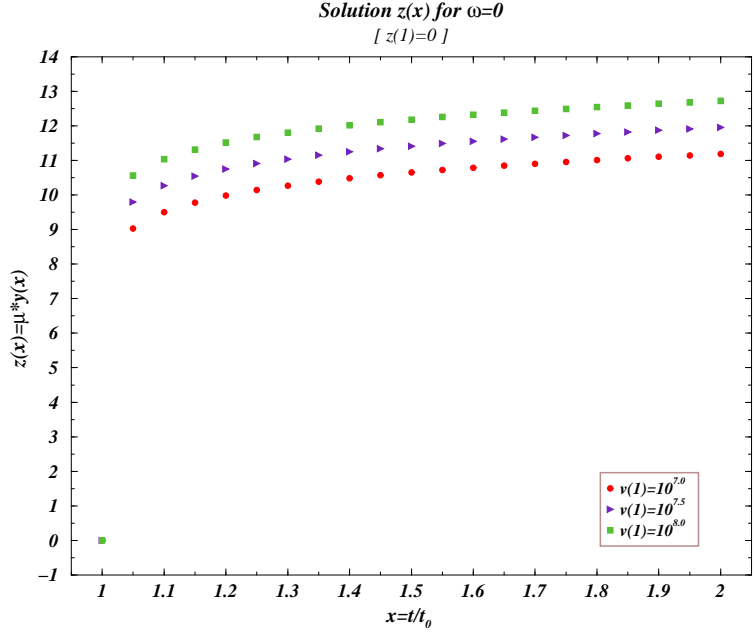}}  \nonumber
\end{eqnarray}
\end{center}
\caption{{Same as before, with positive initial velocity}}
\label{4}
\end{figure}
%%%%%%%%%%%%%%%%%%%%%%% End Figure 4  %%%%%%%%%%%%%%%%%%%%%%%%%%%%%% 

%%%%%%%%%%%%%%%%%%%%%% Begin Figure 5 %%%%%%%%%%%%%%%%%%%%%%%%%%%%%%
\begin{figure}[htb]%[ht]
\begin{center}
\leavevmode       
\begin{eqnarray}
\epsfxsize= 8truecm\rotatebox{-90}{\epsfbox{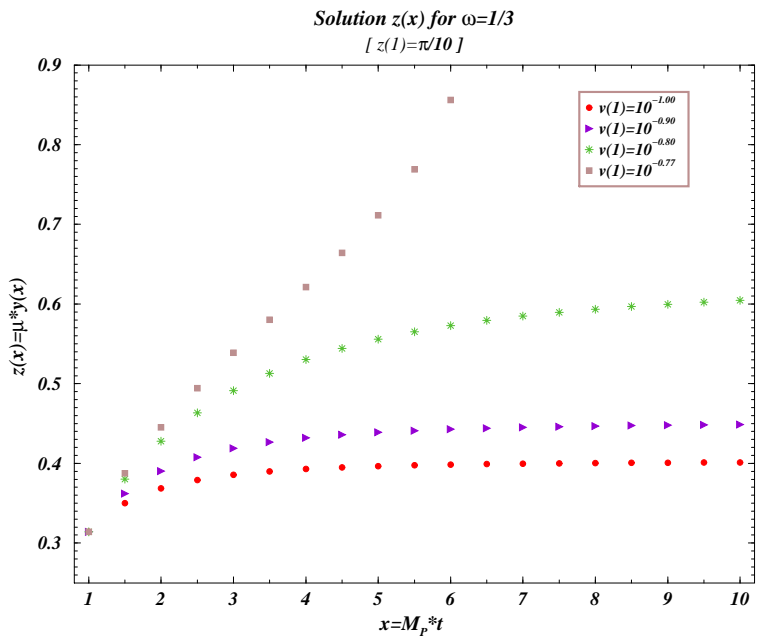}}  \nonumber
\end{eqnarray}
\end{center}
\caption{{Diagram for $y\sim 0.3 \ell_P$ in the radiation dominated
    case. Notice the plateau followed in the case of lowest initial
    velocity.}} 
\label{5}
\end{figure}
%%%%%%%%%%%%%%%%%%%%%%% End Figure 5  %%%%%%%%%%%%%%%%%%%%%%%%%%%%%% 

%%%%%%%%%%%%%%%%%%%%%% Begin Figure 6 %%%%%%%%%%%%%%%%%%%%%%%%%%%%%%
\begin{figure}[htb]%[ht]
\begin{center}
\leavevmode       
\begin{eqnarray}
\epsfxsize= 8truecm\rotatebox{-90}{\epsfbox{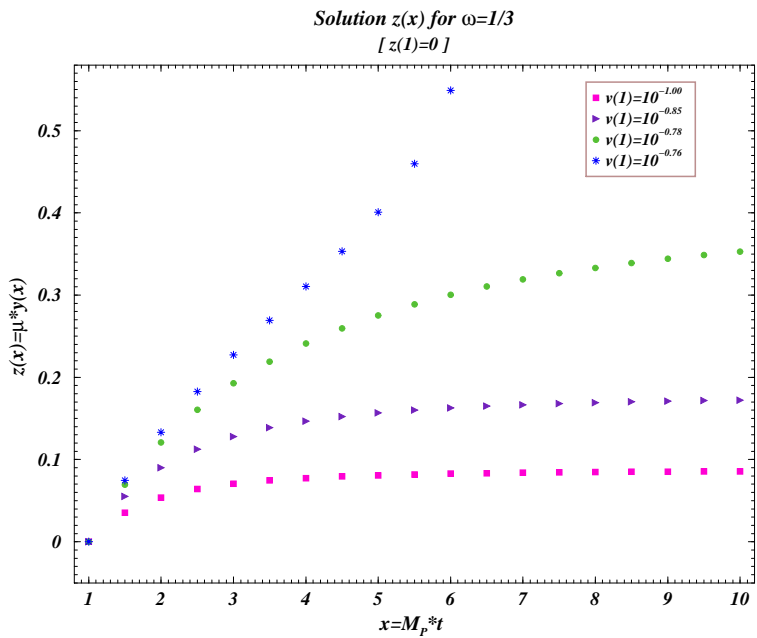}}
\nonumber 
\end{eqnarray}
\end{center}
\caption{{Same as before, with vanishing initial position with respect to
    the brane.}}
\label{6}
\end{figure}
%%%%%%%%%%%%%%%%%%%%%%% End Figure 6  %%%%%%%%%%%%%%%%%%%%%%%%%%%%%% 

%%%%%%%%%%%%%%%%%%%%%% Begin Figure 7 %%%%%%%%%%%%%%%%%%%%%%%%%%%%%%
\begin{figure}[t]%[h]
\begin{center}
\leavevmode       
\begin{eqnarray}
\epsfxsize= 8truecm\rotatebox{-90}{\epsfbox{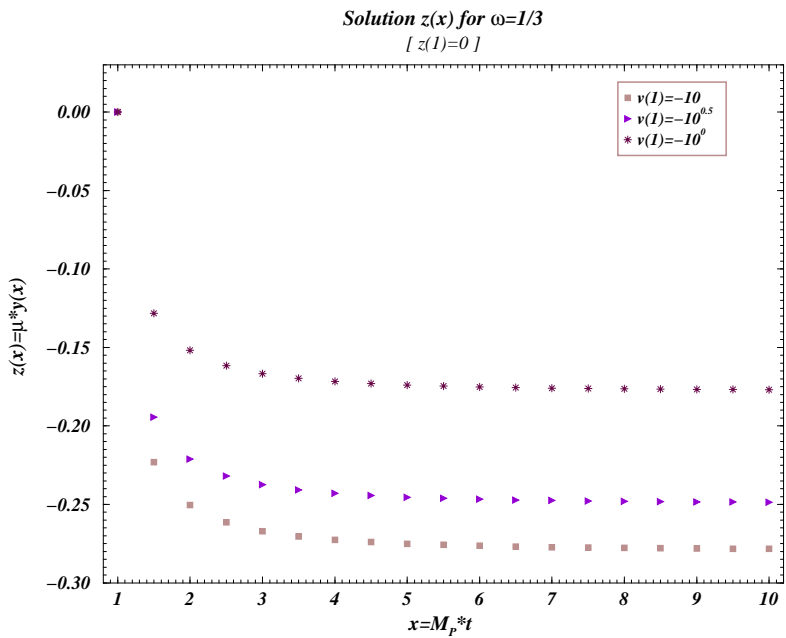}}  \nonumber
\end{eqnarray}
\end{center}
\caption{{Same as before, with negative initial velocity}}
\label{7}
\end{figure}
%%%%%%%%%%%%%%%%%%%%%%% End Figure 7  %%%%%%%%%%%%%%%%%%%%%%%%%%%%%% 

%%%%%%%%%%%%%%%%%%%%%%%%%%%%%%%%%%%%%%%%%%%%%%%%%%%%%%%%%%%%%%%%%%%%%
%%%%%%%%%%%%%%%%%%%%%%%%%%%%%%%%%%%%%%%%%%%%%%%%%%%%%%%%%%%%%%%%%%%%%
%%%%%%%%%%%%%%%%%%%%%%%%%%%%%%%%%%%%%%%%%%%%%%%%%%%%%%%%%%%%%%%%%%%%%

Letting $y=tf(t)$ in (\ref{e-l}), we get a nonlinear differential equation
\begin{eqnarray}
&&-[1+(2q\omega-q)f
+(q^2\omega^2-2q^2\omega^2)f^2-q^3\omega^2f^3](t^2\ddot{f}+2t\dot{f})\nonumber
\\
&&+[q+(2\omega
q^2-q^2+q-q\omega)f\nonumber\\&&+(q^2\omega-q^2\omega^2-2\omega
q^3+2\omega q^2)f^2\nonumber\\
&&+(2q^3\omega^2-q^4\omega^2)f^3](t\dot{f}+f)
\nonumber\\&&+[2q\omega-q)+q^3\omega^2f^2](t\dot{f}+f)^2\nonumber\\&&
-[q-q(q-1)f](t\dot{f}+f)^3\nonumber\\&&
+[(-q-q\omega)
+(q^2-4q^2-3q^2\omega^2)f\nonumber\\&&
+(3q^3\omega-6q^3\omega^2-3q^3\omega^3)f^2\nonumber\\&&
+(-2q^4\omega^3+3q^4-q^4\omega^4)f^3\nonumber\\&&
+(q^5\omega^3-q^5\omega^4)f^4]
=0 \quad .
\end{eqnarray}

The  analysis of such a differential equation is beyond our
capability. We leave it as it stands and pass to a discussion of some
simple cases where numerical analysis can be performed.

The case considered by Bin\'etruy et al. \cite{s15} is that of a
3-brane universe in the 5-dimensional space time with a cosmological
constant. For an equation of state $p =\omega \rho$ they found
explicit solutions which we use in order to study the question of the
existence of shortcuts. The solutions are very involved, and we first
disentangle the equations using a MAPLE program, and further on
numerically solve the differential equations. We shall consider the
matter dominated ($\omega =0$) and radiation dominated ($\omega=1/3$) cases.

The solution of the gravity equations reads \cite{s15}
\begin{eqnarray}\label{binetruy}
a(t,y) = \left\{ {1 \over 2} \left( 1+{{\kappa^2 \rho_b ^2} \over {6
        \rho_B}} \right) + {1 \over 2} \left( 1-{{\kappa^2 \rho_b ^2} \over {6
        \rho_B}} \right) \cosh (\mu y) - {{\kappa \rho_b} \over
        \sqrt{-6 \rho_B}} \sinh (\mu \mid y \mid) \right\}^{1 \over 2} a_0
        (t) \, \, , && \nonumber \\ \\
n(t,y) = {{\dot a (t,y)} \over {\dot a_0 (t)}} \, \, , \qquad \qquad
        \qquad \qquad \qquad \qquad \qquad  \quad && \nonumber
\end{eqnarray}
where 
\begin{eqnarray}
&a_0 (t) = a_\star (\kappa^2 \rho_\star)^{1/q} \left( {{q^2} \over
    {72}} \kappa^2 \rho_\Lambda t^2 + {q \over 6} t \right) ^{1/q}
    \quad , & \nonumber \\ \\
&\mu = \sqrt {-{{2 \kappa^2} \over 3} \rho_B} & \nonumber
\end{eqnarray}
with $a_\star$ and $\rho_\star$ constants.

In addition, $\rho_b$ and $\rho_B$ are the matter densities on the brane
and on the bulk respectively. We have to choose these constants, which we
do according to the course we are using to discuss the possibilities of
shortcuts. We choose the parameters according to the discussion in
Bin\'etruy et al. \cite{s7}
\be
\rho_b=\rho_\Lambda + \rho \quad , 
\ee
where $\rho$ stands for the ordinary energy density in cosmology given
by
\be
\rho=\rho_\star (a_0/a_\star)^{-q} \quad , \qquad q=3(1+\omega) \quad .
\ee

The intrinsic tension of the brane, $\rho_\Lambda$, has to be identified with
Newton's constant in order to recover the standard cosmology, that is
\be
8\pi G = {{\kappa^4 \rho_\Lambda} \over 6} \quad ,
\ee
when $\rho \ll \rho_\Lambda$.

Moreover the 5 dimensional coupling constant $\kappa$,
the 5-dimensional Newton constant $G_{(5)}$, and the Planck mass
$M_{(5)}$ are related by 
\be
\kappa^2 = 8 \pi G_{(5)} = M^{-3} _{(5)} \quad .
\ee

Furthermore, we follow Randall and Sundrum and relate the bulk energy
density $\rho_B$ and the cosmological constant density $\rho_\Lambda$
by
\be
\rho_B = -{{\kappa^2 \rho_\Lambda^2} \over 6} \quad .
\ee

At this point all constants are defined in terms of the Planck mass,
and our discussion of the evolution of gravity signs can be established.

%%%%%%%%%%%%%%%%%%%%%%%%%%%%%%%%%%%%%%%%%%%%%%%%%%%%%%%%%%%%%%%%%%%%%
%%%%%%%%%%%%%%%%%%%%%%%%%%%%%%%%%%%%%%%%%%%%%%%%%%%%%%%%%%%%%%%%%%%%%
For the matter dominated case, $\omega =0$, we experimented using
different initial conditions. In general, we prefer to start with $y\not =
0$ in order to avoid any spurious solution in the differential equation,
which is rather singular. We thus suppose that $y$ starts at the order of
the Planck length. Pictures \ref{2} to \ref{4} show some results. We
have chosen to plot the adimensional function $z(x)=\mu y(x)$, where
$\mu$ corresponds to twice Planck mass units $M_P$ and $x=t/t_0$, $t_0$
being the present age of the universe.

Each graph contains a set of curves corresponding to three typical
velocities, whose values are shown in the legend of each graph,
producing similar behaviors. In figures \ref{2} and \ref{3} we use
negative initial velocities and, independently of the chosen initial
point $y$, the curve decays and escapes, never returning to the same brane. 
In the case of positive initial velocities, picture \ref{4} shows
three curves from which we can notice that the greater initial
velocity is, the further away from the brane the object will travel.

Summarizing, these graphs show that the gravity wave always ``tries to
follow the brane'', since the $y$ coordinate either drops fast to
zero, or drives away, which means that the final point reached is far from the original brane.

We thus conjecture, based on these results, that the shortest path is
inside the brane, being the one followed by light. However, there is
certainly room for further paths due to the extremely complicated
character of the differential equation involved in the problem. Moreover,
there seems to be some attractors in the differential equation, which
further complicate the matter, rendering a possible solution even more
obscure, while opening further possibilities of shortcuts, especially in
cases where the bulk density becomes important.

Such complications actually do not arise in full in the matter dominated
case, but can be clearly seen in the radiation dominated era. In these cases,
solutions are shown in figures \ref{5} through \ref{7}. Again, we have
plotted the adimensional function $z(x)$, where $x=M_P \,\, t$ in this
case.

Pictures \ref{5} and \ref{6} show a plateau behavior for low positive initial
velocities; however, there is a threshold velocity for which the curve
decouples and escapes to infinity. Picture \ref{7} shows curves
for three negative initial velocities. Again, the wave tries to follow
the brane from a distance depending on the initial velocity value as
we had seen in matter dominated case.

In the radiation dominated era, $\omega =\frac 13$, attractors are more
clearly formed. Their meaning is not known and in some cases, where we can
avoid dropping into them using special initial conditions, it is natural
to foresee solutions which return to the brane after a roundabout in the
bulk, although we have to stress that no such solution has been found so
far.  We leave  this more difficult numerical problem for a future 
publication.

{\bf Acknowledgements:} This work was supported  by
Funda\c c\~ao de Amparo \`a Pesquisa do Estado de
S\~ao Paulo (FAPESP) and Conselho Nacional de Desenvolvimento
Cient\'\i fico e Tecnol\'ogico (CNPq), Brazil, and
NNSF of China.\\
e-mails: eabdalla@fma.if.usp.br; bertha@fma.if.usp.br;
sshfeng@fma.if.usp.br; binwang@fma.if.usp.br;

%%%%%%%%%%%%%%%%%%%%%%%%%%%%%%%%%%%%%%%%%%%%%%%%%%%%%%%%%%%%%%%%%%%%%%

\end{document}